\documentclass[11pt,a4paper]{article}

\newif\ifjhepstyle
\jhepstylefalse
\jhepstyletrue

\ifjhepstyle
	\usepackage{jheppub}
	\usepackage{amsfonts}
	\usepackage{verbatim}
	\usepackage{float}
	\restylefloat{table}
	
\else	\usepackage{verbatim}
	\usepackage{cite}
	\usepackage{setspace}
	\usepackage[top=2.45cm, bottom=2.5cm, left=2.45cm, right=2.45cm]{geometry}
	\usepackage{amsmath,amsfonts,amssymb}
	\usepackage[colorlinks=true
	,urlcolor=blue
	,anchorcolor=blue
	,citecolor=blue
	,filecolor=blue
	,linkcolor=blue
	,menucolor=blue
	]{hyperref}
	\usepackage{float}
	\restylefloat{table}
	
	\numberwithin{equation}{section}
\fi

\newcommand{\ThisIsTheTitle}{Strings from 3D gravity: asymptotic dynamics of AdS$_3$ gravity with free boundary conditions}
\newcommand{\ThisIsAuthorOne}{Luis Apolo}
\newcommand{\ThisIsEmailOne}{luis.apolo@fysik.su.se}
\newcommand{\ThisIsAuthorTwo}{Bo Sundborg}
\newcommand{\ThisIsEmailTwo}{bo@fysik.su.se}
\newcommand{\ThisIsTheAffiliation}{Department of Physics \& The Oskar Klein Centre, \\
Stockholm University, AlbaNova University Centre, SE-106 91 Stockholm, Sweden}
\newcommand{\TheseAreTheKeywords}{Classical Theories of Gravity, Bosonic Strings, Chern-Simons Theories}

\newcommand{\ThisIsTheAbstract}{Pure three-dimensional gravity in anti-de Sitter space can be formulated as an SL(2,R) $\times $ SL(2,R) Chern-Simons theory, and the latter can be reduced to a WZW theory at the boundary. In this paper we show that AdS$_3$ gravity with free boundary conditions is described by a string at the boundary whose target spacetime is also AdS$_3$. While boundary conditions in the standard construction of Coussaert, Henneaux, and van Driel are enforced through constraints on the WZW currents, we find that free boundary conditions are partially enforced through the string Virasoro constraints.}

\ifjhepstyle
\title{\ThisIsTheTitle}

\author{\ThisIsAuthorOne}
\author{and \ThisIsAuthorTwo}

\affiliation{\ThisIsTheAffiliation}

\emailAdd{\ThisIsEmailOne}
\emailAdd{\ThisIsEmailTwo}

\abstract{\ThisIsTheAbstract} 

\fi

\begin{document}

\ifjhepstyle
\maketitle
\flushbottom
\fi

\long\def\symfootnote[#1]#2{\begingroup%
\def\thefootnote{\fnsymbol{footnote}}\footnote[#1]{#2}\endgroup} 

\def\({\left (}
\def\){\right )}
\def\lb{\left [}
\def\rb{\right ]}
\def\lB{\left \{}
\def\rB{\right \}}

\def\Int#1#2{\int \textrm{d}^{#1} x \sqrt{|#2|}}
\def\Bra#1{\left\langle#1\right|} 
\def\Ket#1{\left|#1\right\rangle}
\def\BraKet#1#2{\left\langle#1|#2\right\rangle} 
\def\Vev#1{\left\langle#1\right\rangle}
\def\Vevm#1{\left\langle \Phi |#1| \Phi \right\rangle}\def\bbox{\bar{\Box}}
\def\til#1{\tilde{#1}}
\def\wtil#1{\widetilde{#1}}
\def\ph#1{\phantom{#1}}

\def\ra{\rightarrow}
\def\la{\leftarrow}
\def\lra{\leftrightarrow}
\def\p{\partial}
\def\diff{\mathrm{d}}

\def\sinh{\mathrm{sinh}}
\def\cosh{\mathrm{cosh}}
\def\tanh{\mathrm{tanh}}
\def\coth{\mathrm{coth}}
\def\sech{\mathrm{sech}}
\def\csch{\mathrm{csch}}

\def\a{\alpha}
\def\b{\beta}
\def\g{\gamma}
\def\d{\delta}
\def\e{\epsilon}
\def\ve{\varepsilon}
\def\k{\kappa}
\def\l{\lambda}
\def\n{\nabla}
\def\om{\omega}
\def\s{\sigma}
\def\t{\theta}
\def\z{\zeta}
\def\vp{\varphi}

\def\ss{\Sigma}
\def\dd{\Delta}
\def\gg{\Gamma}
\def\ll{\Lambda}
\def\tt{\Theta}

\def\D{{\cal D}}
\def\F{{\cal F}}
\def\H{{\cal H}}
\def\I{{\cal I}}
\def\J{{\cal J}}
\def\K{{\cal K}}
\def\L{{\cal L}}
\def\O{{\cal O}}
\def\P{{\cal P}}
\def\W{{\cal W}}
\def\X{{\cal X}}
\def\Z{{\cal Z}}

\def\we{\wedge}

\def\zz{\bar z}
\def\xx{\bar x}
\def\xp{x^{+}}
\def\xm{x^{-}}

\def\VirU1{\mathrm{Vir}\otimes\hat{\mathrm{U}}(1)}
\def\VirSL2R{\mathrm{Vir}\otimes\widehat{\mathrm{SL}}(2,\mathbb{R})}
\def\U1{\hat{\mathrm{U}}(1)}
\def\SL2R{\widehat{\mathrm{SL}}(2,\mathbb{R})}
\def\sl2r{\mathrm{SL}(2,\mathbb{R})}
\def\by{\mathrm{BY}}

\def\tr{\mathrm{tr}}

\def\sint{\int_{\ss}}
\def\dsint{\int_{\p\ss}}

\newcommand{\eq}[1]{\begin{align}#1\end{align}}
\newcommand{\eqst}[1]{\begin{align*}#1\end{align*}}

\newcommand{\absq}[1]{{\textstyle\sqrt{|#1|}}}


\unless\ifjhepstyle
\begin{titlepage}
\begin{center}

\ph{.}

\vskip 4 cm

{\Large \bf \ThisIsTheTitle}

\vskip 1 cm

{{\ThisIsAuthorOne} and {\ThisIsAuthorTwo}}

\vskip .75 cm

{\em \ThisIsTheAffiliation}

\end{center}

\vskip 1.25 cm

\begin{abstract}
\noindent \ThisIsTheAbstract
\end{abstract}
\end{titlepage}
\newpage
\fi

\unless\ifjhepstyle
\tableofcontents
\noindent\hrulefill
\bigskip
\fi


\section{Introduction}

Three-dimensional gravity~\cite{Deser:1983tn,Deser:1983nh} is a rich system wherein to study aspects of quantum gravity and holography. Despite having no local degrees of freedom, the theory possesses boundary gravitons, linearized perturbations localized at the boundary. These can be interpreted as Goldstone bosons of the spontaneously broken Virasoro symmetries found by Brown and Henneaux~\cite{Brown:1986nw}. Although all solutions to Einstein's equations are locally flat, or (anti) de Sitter in the presence of a cosmological constant, the theory contains globally non-trivial solutions. In the case of a negative cosmological constant these solutions are BTZ black holes~\cite{Banados:1992wn, Banados:1992gq}.

The existence of black holes in AdS$_3$ should play an important role in the quantum theory. The latter, according to the to the AdS/CFT correspondence~\cite{Maldacena:1997re,Gubser:1998bc,Witten:1998qj}, is a two-dimensional conformal field theory~\cite{Witten:2007kt}. An attempt to make sense of these statements for pure gravity, described only by the Einstein-Hilbert action without additional matter, was made by Maloney and Witten~\cite{Maloney:2007ud}. There, the partition function of the Euclidean theory was computed and found to be inconsistent with that of a two-dimensional CFT. Many assumptions were made in this calculation; in particular, that the spectrum of AdS$_3$ gravity consists only of BTZ black holes and their SL($2,\mathbb{Z}$) cousins~\cite{Maldacena:1998bw}. Thus, it seems that if pure gravity in AdS$_3$ is to make sense as a quantum theory we need to consider additional elements (or reconsider other assumptions)~\cite{Maloney:2007ud,Keller:2014xba}. Steps towards this direction have been taken recently in~\cite{Kim:2014bga,Kim:2015bba} where long strings were added to the spectrum. It was also observed a long time ago that the Chern-Simons description of 3D gravity~\cite{Achucarro:1987vz,Witten:1988hc} contains classical solutions missing in the metric description~\cite{Mansson:2000sj}. Such solutions play an important and still mysterious role in higher spin extensions of 3D gravity~\cite{Castro:2011iw}.

Even if we do not know what exactly is missing in the calculation of ref.~\cite{Maloney:2007ud}, we can derive the entropy of BTZ black holes by making reasonable assumptions about the quantum theory -- i.e.~about the dual CFT. Cardy's formula on the asymptotic growth of states~\cite{Cardy:1986ie} and the central charge of the Virasoro algebra computed by Brown and Henneaux~\cite{Brown:1986nw} are enough ingredients to reproduce the black hole entropy~\cite{Strominger:1997eq}. Although there have been many attempts to derive the entropy of black holes from 3D gravity itself (see~\cite{Carlip:1995qv} for a review) a satisfying answer, e.g.~a concrete and calculable 2D CFT description, is still lacking. Indeed, to compute or even estimate the entropy of black holes without appealing to Cardy's formula it is necessary to have a deeper understanding of the spectrum of 3D gravity. 


To address the above problems we propose an alternative formulation of 3D gravity wherein to analyze the spectrum of the theory. We consider a variation of the standard Brown-Henneaux boundary conditions, known as free boundary conditions, where the boundary metric is allowed to fluctuate~\cite{Compere:2008us,Compere:2013bya,Troessaert:2013fma,Avery:2013dja,Apolo:2014tua}. Since three-dimensional gravity has no local degrees of freedom propagating in the bulk this is a minimal modification of the pure theory. We will show that the asymptotic dynamics of the theory are described by a string whose target spacetime is also AdS$_3$. This is achieved by finding a Chern-Simons formulation of 3D gravity that is analytic in the gauge fields and compatible with free boundary conditions. An important step towards this formulation is a careful analysis of boundary terms and boundary conditions. The essential new ingredient is a reformulation of the area term\footnote{This is the boundary counterterm that renders the action of AdS$_3$ gravity finite~\cite{Balasubramanian:1999re}. We will see below that in 3D gravity it is possible to change the coefficient of the Gibbons-Hawking term. Then the area term is not only necesssary to regularize the action but also to keep the variational principle well-defined.} regularizing the action of 3D gravity which will be seen to correspond to a diffeomorphism-invariant boundary term in the Chern-Simons theory.\footnote{Alternatively, we could have started with diffeomorphism-invariant boundary terms in the Chern-Simons theory and required that they lead to a well-defined action in the metric theory.} The latter is necessary in our approach since free boundary conditions in the bulk are analogous to a diffeomorphism-invariant description of the asymptotic boundary -- this is the main principle behind our results.

Crucial related observations were made by Arcioni, Blau, and O'Loughlin in~\cite{Arcioni:2002vv} but their construction does not imply the Virasoro constraints we obtain for the boundary theory. Several aspects of string theory on AdS$_3$ have been studied in the literature, see e.g.~\cite{Balog:1988jb,Hwang:1990aq,Henningson:1991jc,Evans:1998qu,Giveon:1998ns,deBoer:1998pp,Kutasov:1999xu,Maldacena:2000hw,Maldacena:2000kv,Maldacena:2001km} and references therein, and these in turn may have important implications for pure three-dimensional gravity.
 
In the standard construction of Coussaert, Henneaux, and van Driel~\cite{Coussaert:1995zp} the WZW model describing the asymptotic dynamics of 3D gravity is constrained and reduces to a Liouville theory. This is because the currents of the WZW model are proportional to the vielbeins and spin connection of the first order formulation of gravity and these must satisfy the Brown-Henneaux boundary conditions~\cite{Brown:1986nw}. In contrast, we find that the SL(2,R) currents of the string are not proportional to the vielbeins and spin connection of 3D gravity. This means that free boundary conditions do not lead to constraints on the SL(2,R) currents and there is no automatic reduction of the WZW model into a Liouville theory. Instead, it is the Virasoro constraints that render the spectrum of the theory unitary~\cite{Hwang:1990aq,Evans:1998qu,Maldacena:2000hw}. This is crucial since the SL(2,R) WZW theory contains states of negative norm.

The present work is a realization of the ideas proposed in~\cite{Banados:1999gh,Sundborg:2013bya} where it was argued that solutions to pure three-dimensional gravity can be mapped into solutions of string theory. There it was shown that some of the constraints on the SL(2,R) currents that enforce Brown-Henneaux boundary conditions can be replaced by Virasoro constraints. Here we find that the Virasoro constraints are a result of imposing free boundary conditions. An analysis of string solutions and their consequences for three-dimensional gravity will be presented elsewhere.

The paper is organized as follows. In Section~\ref{se:generalizedgraction} we describe free boundary conditions and reconsider the boundary terms in the action of AdS$_3$ gravity. In Section~\ref{se:cstowzw} we obtain a first-order (Chern-Simons) formulation of 3D gravity compatible with free boundary conditions. We also discuss its reduction to a WZW/string action at the boundary and show that free boundary conditions are not enforced through constraints linear on the currents of the theory. We end with a discussion and conclusions in Section~\ref{se:conclusions}.


\section{Generalized gravitational action and free boundary conditions} 
\label{se:generalizedgraction}

In this section we introduce free boundary conditions and argue that boundary terms in the Chern-Simons formulation of the theory must be generally covariant. We also consider a generalized gravitational action differing from the standard Einstein-Hilbert action by the choice of boundary terms. In particular, we will show that this action is well-defined when free boundary conditions hold. The Chern-Simons formulation of AdS$_3$ gravity with free boundary conditions is constructed directly from the first order formulation of this generalized gravitational action in the next section.

\subsection{Free boundary conditions} \label{suse:freebc}

Let us begin by discussing free boundary conditions, an alternative to Brown-Henneaux boundary conditions where the boundary metric becomes dynamical~\cite{Compere:2008us}. The boundary conditions corresponding to a boundary metric in the lightcone and conformal gauges were studied in refs.~\cite{Compere:2013bya,Troessaert:2013fma,Avery:2013dja,Apolo:2014tua}. We can generalize their results and assume that, for any gauge, free boundary conditions are given by
 \eq{
  g_{\mu\nu} & = r^2 h^{(0)}_{\mu\nu} + \O(r^0), \hspace{30pt} g_{r \mu} = \O(r^{-3}), \hspace{30pt} g_{rr} = r^{-2} + \O(r^{-4}),  \label{se2:freebc}
  }
where greek indices denote boundary coordinates and $h^{(0)}_{\mu\nu}$ is the metric at the boundary ($r \ra \infty$).\footnote{Nothing prevents us from having stronger boundary conditions but these are most likely uninteresting. The boundary conditions in eq.~\eqref{se2:freebc} are physically interesting since they allow for solutions with finite, non-zero charges such as mass and angular momentum.} Thus, free boundary conditions are a straightforward generalization of Brown-Henneaux boundary conditions~\cite{Brown:1986nw}, the only difference being that $h^{(0)}_{\mu\nu}$ is a dynamical object in the former, a background metric in the latter. This seemingly innocuous modification of the boundary conditions has important consequences -- it changes the asymptotic symmetries of the theory~\cite{Compere:2013bya,Troessaert:2013fma,Avery:2013dja}. 

Note that in refs.~\cite{Avery:2013dja,Apolo:2014tua} weaker boundary conditions were assumed when the boundary metric is in the lightcone gauge, specifically $g_{r \hat{\mu}} = \O(r^{-1})$ where $x^{\hat{\mu}}$ is \emph{one} of the boundary coordinates. However, it is possible to strengthen the boundary conditions so as to agree with~\eqref{se2:freebc} and still keep the $\widehat{SL}$(2,R) symmetries found in~\cite{Avery:2013dja} intact. In fact, although this was not noticed in~\cite{Apolo:2014tua}, the \emph{improved} $\widehat{SL}$(2,R) generators introduced there accomplish just that. Thus, for all known cases the above boundary conditions lead to well-behaved theories with finite, conserved, and integrable charges.

An important aspect of free boundary conditions is that they do not come for free. In order to have a well-defined variational principle the Brown-York stress-energy tensor must vanish~\cite{Compere:2008us}, as we now demonstrate. Indeed, when the gravitational action is given by the standard expression,
  \eq{ 
  S = \frac{1}{\k} \left \{ \frac{1}{2} \sint \absq{g} \(R + \frac{2}{\ell^2} \) + \dsint\absq{h} K - \frac{1}{\ell} \dsint\absq{h} \right \}, \label{se2:graction}
  }
where $\k = 8 \pi G_N$, $G_N$ is Newton's constant, $\ell$ is the radius of AdS which we set to one, $h$ is the determinant of the induced metric at the boundary, and $K$ is the trace of the extrinsic curvature defined in~\eqref{se2:extrinsic}; variation of the action yields
  \eq{
  \d S = \frac{1}{2\k} \sint \absq{g} \Pi_{\mu\nu} \d g^{\mu\nu} -\frac{1}{2} \dsint \absq{h^{(0)}} T^{BY}_{\mu\nu} \d h^{(0)\mu\nu}. \label{se2:variation0}
  }
Here $0 = \Pi_{\mu\nu} = R_{\mu\nu} - \frac{1}{2} R g_{\mu\nu} - g_{\mu\nu}$ are the bulk equations of motion, $h^{(0)}_{\mu\nu}$ is the boundary metric, and $T_{\mu\nu}^{BY}$ is the Brown-York stress-energy tensor given by~\cite{Brown:1992br,Balasubramanian:1999re} \footnote{Note that our definition of the extrinsic curvature in eq.~\eqref{se2:extrinsic} differs by a sign with respect to that in ref.~\cite{Balasubramanian:1999re}.}
  \eq{
  T^{BY}_{\mu\nu} = -\frac{2}{\absq{h^{(0)}}} \frac{\d S}{\d h^{(0)\mu\nu}} = -\frac{1}{\k} \big ( K_{\mu\nu} - K h_{\mu\nu} + h_{\mu\nu} \big ). \label{se2:bytmunu}
  }
Thus, to recover a well-defined variational principle we demand that the Brown-York tensor vanishes\footnote{Alternatively one can add boundary terms to the action that cancel the on-shell value of the Brown-York tensor~\cite{Compere:2013bya,Avery:2013dja}. Also note that one can interpret the vanishing of the Brown-York tensor as a boundary condition instead of an equation of motion~\cite{Compere:2008us,Troessaert:2013fma}.}
  \eq{
  T^{BY}_{\mu\nu} = 0.  \label{se2:bytmunuzero}
  }
This equation can be interpreted as the equation of motion of the now dynamical boundary metric. In three-dimensions eq.~\eqref{se2:bytmunuzero} is compatible with the equations of motion obtained from the conformal anomaly~\cite{Apolo:2014tua}.


\subsection{Generalized gravitational action} \label{suse:generalizedgraction}

We would like to incorporate free boundary conditions in the Chern-Simons formulation of AdS$_3$ gravity. Once we upgrade from Brown-Henneaux boundary conditions we face an immediate challenge. Since the boundary metric is now dynamical, boundary terms in the Chern-Simons formulation of the theory should be generally covariant. In other words, we will demand full reparametrization invariance of the asymptotic boundary. When the boundary metric is not dynamical this is not essential~\cite{Coussaert:1995zp}. One way to find the appropriate boundary terms is to construct the vielbein counterparts to the terms appearing in the Einstein-Hilbert action~\eqref{se2:graction}. It turns out that the coefficients in front of these boundary terms are not unique, however. 

Let us therefore consider the generalized gravitational action studied in ref.~\cite{Detournay:2014fva}
  \eq{
  S_{GR} = \frac{1}{\k} \left \{ \frac{1}{2} \sint \absq{g} \(R + 2\) + \a \dsint \absq{h} K + \b \dsint \absq{h} \right \}. \label{se2:generalizedgraction}
  }
The authors of~\cite{Detournay:2014fva} were interested in the flat space limit of eq.~\eqref{se2:graction}. This limit is not well-defined since the area term, i.e.~the last term, in eq.~\eqref{se2:graction} contains a factor of $\ell^{-1}$. On the other hand if the generalized action~\eqref{se2:generalizedgraction} is well-defined in the limit $\b \ra 0$, taking the $\ell \ra 0$ limit is not a problem. This is indeed the case if $\a = 1/2$~\cite{Miskovic:2006tm}. More generally, the action~\eqref{se2:generalizedgraction} is valid -- it yields a well-defined variational principle and a finite Brown-York stress tensor -- for any $\a$ and $\b$ satisfying $2\a + \b = 1$ provided that Brown-Henneaux boundary conditions hold~\cite{Detournay:2014fva}. 

Let us show that the results of ref.~\cite{Detournay:2014fva} hold for free boundary conditions as well. We will see that the action has a well-defined variational principle, so that variation of the action yields eq.~\eqref{se2:variation0}, and that the corresponding Brown-York tensor is finite and given by eq.~\eqref{se2:bytmunu} \emph{on-shell}.

A parametrization of the metric compatible with the boundary conditions is given by
  \begin{equation}
  \begin{split}
  g_{\mu\nu} & = r^2 h^{(0)}_{\mu\nu} + h^{(2)}_{\mu\nu} + \O(r^{-2}), \\
  g_{rr} & = r^{-2} + r^{-4} g^{(2)}_{rr} + \O(r^{-6}), \\
  g_{r\mu} & = \O(r^{-3}),
  \end{split} \label{se2:bcgauge}
  \end{equation}
where $h^{(j)}_{\mu\nu}$ are components of the induced metric at the boundary,
  \eq{
  h_{\mu\nu} = r^{2} h^{(0)}_{\mu\nu} + h^{(2)}_{\mu\nu} + \sum_{j=1} r^{-2j} h^{2(1+j)}_{\mu\nu}.\label{se2:inducedmetric}
  } 
With this parametrization, the (normalized, outward pointing) vector normal to the boundary is given by
  \eq{
  n^{M} = \left [ r - \frac{1}{2r} g^{(2)}_{rr} + \O(r^{-3}) \right ] \d^{M}_{r},  \label{se2:normalvector}
  }
where uppercase Roman indices denote bulk spacetime coordinates. The fall-off condition on the $g_{r\mu}$ components of the metric given in eq.~\eqref{se2:freebc} is important. Had we chosen $g_{r\mu} = \O(r^{-1})$ instead, the dual normal vector $n_{M}$ would receive corrections of order 1 that spoil the results presented below.\footnote{Note, however, that in our analysis $h^{(0)}_{\mu\nu}$ is arbitrary. It is still possible that for particular choices of $h^{(0)}_{\mu\nu}$ weaker boundary conditions on $g_{r\mu}$ lead to a well-defined action.} It is thus comforting that eq.~\eqref{se2:freebc} holds for the free boundary conditions studied in the literature. Finally let us state our conventions for the extrinsic curvature, which carries the opposite sign with respect to ref.~\cite{Balasubramanian:1999re}
  \eq{
  K_{MN} = \nabla_{(M}n_{N)} = \frac{1}{2} \( \nabla_{M} n_{N} + \nabla_{N} n_{M} \).  \label{se2:extrinsic}
  }
The trace of the extrinsic curvature is computed using the induced metric, i.e.~$K = h^{\mu\nu} K_{\mu\nu}$. Note that we raise indices using the induced metric since the $g_{r\mu}$ components of the metric scale as $r^{-3}$ and drop off from the boundary terms in the action.

The variation of the generalized action~\eqref{se2:generalizedgraction} is given by~\cite{Detournay:2014fva}
  \eqst{
  \d \(2\k\,S_{GR}\) & = \sint \absq{g} \,\Pi_{\mu\nu} \d g^{\mu\nu} +  \dsint \absq{h} \big [ K_{\mu\nu} - \( \a K + \b \) h_{\mu\nu} \big ] \d g^{\mu\nu} \\ 
  & + \dsint \absq{h} \(\a -1 \) \big ( K  n_{\mu} n_{\nu} - h_{\mu\nu} n^{\rho} \nabla_{\rho} \big ) \d g^{\mu\nu}  ,
  }
where $\Pi_{\mu\nu} = 0$ are the bulk equations of motion and we have omitted terms characteristic of non-smooth boundaries (see~\cite{Detournay:2014fva} and references therein). Using eqs.~\eqref{se2:bcgauge},~\eqref{se2:normalvector}, and~\eqref{se2:extrinsic} we obtain
  \eq{
   \d S_{GR} & =  \frac{1}{2\k} \sint \absq{g}\,\Pi_{\mu\nu} \d g^{\mu\nu} - \frac{1}{2\k} \dsint \absq{h^{(0)}} \, \d{\cal I} \label{se2:variation2}
  }
where $\d\I$ is given by
  \begin{equation}
  \begin{split}
  \d \I = & \bigg \{ \( 2\a + \b - 1 \) r^2 h^{(0)}_{\mu\nu} + \Big [ \(1 - 2\a \) \frac{1}{2}\, g^{(2)}_{rr}- \a\, h^{(0)\rho\s} h^{(2)}_{\rho\s} \Big ] h^{(0)}_{\mu\nu} - \b h^{(2)}_{\mu\nu} \bigg \} \d h^{(0)\mu\nu} \\
   &+ 2\( \a - 1\) \d g^{(2)}_{rr} - \(\b+1\) h^{(0)\mu\nu} \d h^{(2)}_{\mu\nu}. 
   \end{split} \label{se2:variation3}
   \end{equation}
The $\{\dots\}$ term is proportional to the Brown-York stress tensor $\widetilde{T}^{BY}_{\mu\nu}$ of the generalized gravitational action. Thus, while the stress-tensor is finite whenever $2\a + \b = 1$, the generalized action is seemingly not well-defined for generic $\a$, $\b$ -- the second line vanishes only for $\a = -\b = 1$.

The action~\eqref{se2:generalizedgraction} \emph{does} have a well-defined variational principle however. As noted in~\cite{Detournay:2014fva}, the trick is to use the bulk equations of motion. Here we will proceed in a slightly different way and instead of using the linearized equations of motion resulting from $\Pi_{\mu\nu} = 0$, we will use the holographic Weyl anomaly~\cite{Henningson:1998gx,Skenderis:1999nb}. This is the observation that the trace of the Brown-York tensor defined in eq.~\eqref{se2:bytmunu} reproduces the conformal anomaly of the dual conformal field theory\footnote{Note that the corresponding expression in~\cite{Henningson:1998gx,Skenderis:1999nb} differs from ours by a minus sign. This is a result of the conventions used there, where the Riemann tensor carries an extra minus sign with respect to ours.} 
  \eq{
  T^{BY\mu}_{\phantom{BY\mu}\mu} = \frac{1}{2\k} R^{(0)}. \label{se2:anomaly}
  }
In this equation $R^{(0)}$ is the Ricci scalar associated with the boundary metric $h^{(0)}_{\mu\nu}$. Note that in order to prove eq.~\eqref{se2:anomaly} one uses the bulk equations of motion so our derivation, like that of ref.~\cite{Detournay:2014fva}, is valid on-shell. Using eq.~\eqref{se2:bytmunu} and the parametrization of the metric given in eq.~\eqref{se2:bcgauge} we find
  \eq{
  g^{(2)}_{rr} + h^{(0)\mu\nu} h^{(2)}_{\mu\nu} = -\frac{1}{2} R^{(0)}. \label{se2:anomalybcgauge}
  }
This equation relates the variation of leading and subleading components of the metric, so that we may write
  \eq{
  \d g^{(2)}_{rr} = -\frac{1}{2} \d R^{(0)} - h^{(2)}_{\mu\nu} \d h^{(0)\mu\nu} - h^{(0)\mu\nu} \d h^{(2)}_{\mu\nu}. \label{se2:anomalyvar}
  }
Using eqs.~\eqref{se2:anomalybcgauge} and~\eqref{se2:anomalyvar} in $\d\I$ we finally obtain
  \eq{
  \dsint \absq{h^{(0)}}\d\I = & -\dsint \absq{h^{(0)}} \left [ \( \frac{1}{2} g^{(2)}_{rr} + h^{(0)\rho\s} h^{(2)}_{\rho\s} \) h^{(0)}_{\mu\nu} - h^{(2)}_{\mu\nu} \right ] \d h^{(0)\mu\nu} \notag \\
  & + \(2\a +\b -1\) \dsint \absq{h^{(0)}} \left [ \( r^2 h^{(0)}_{\mu\nu} - h^{(2)}_{\mu\nu} \) \d h^{(0)\mu\nu} - h^{(0)\mu\nu} \d h^{(2)}_{\mu\nu} \right ] \notag \\
  & - \( \a - 1\) \dsint \d \Big ( \absq{h^{(0)}} R^{(0)} \Big ). \label{se2:variationfinal}
  }

The $[...]$ term in the first line of eq.~\eqref{se2:variationfinal} is proportional to the Brown-York stress tensor of the generalized gravitational action. The second line, which was partly responsible for spoiling the variational principle in eq.~\eqref{se2:variation3}, vanishes when $2\a + \b = 1$. Finally, the third line is a total derivative in two dimensions and therefore vanishes. Thus, we conclude that, as long as (1) $2\a + \b = 1$, (2) Brown-Henneaux or free boundary conditions hold, and (3) we are allowed to use the bulk equations on motion in eq.~\eqref{se2:variation3}, variation of the action yields
  \eq{
  \d S_{GR} = \frac{1}{2\k} \sint \absq{g}\, \Pi_{\mu\nu} \d g^{\mu\nu} -\frac{1}{2} \dsint\absq{ h^{(0)}}\, \widetilde{T}^{BY}_{\mu\nu} \d h^{(0)\mu\nu}. \label{se2:variation}
  }
The new Brown-York stress tensor $\widetilde{T}^{BY}_{\mu\nu}$ is finite and independent of $\a$ and $\b$. In fact it is equal to the on-shell value of the standard expression for the Brown-York tensor given in eq.~\eqref{se2:bytmunu}. Using the parametrization of the metric given in eq.~\eqref{se2:bcgauge} we indeed find 
  \eq{
  \widetilde{T}^{BY}_{\mu\nu} = T^{BY}_{\mu\nu} =  -\frac{1}{\k} \( \frac{1}{2} g^{(2)}_{rr} h^{(0)}_{\mu\nu} + h^{(0)\rho\s} h^{(2)}_{\rho\s} h^{(0)}_{\mu\nu} - h^{(2)}_{\mu\nu}  \). \label{se2:bytmunuonshell}
  }

Thus the boundary terms in the action of AdS$_3$ gravity are not unique -- any choice of $\a$ and $\b$ with $2\a + \b = 1$ is as good as any other. This freedom on the choice of boundary terms translates straightforwardly to the Chern-Simons theory. Unfortunately the area term in eq.~\eqref{se2:generalizedgraction} is not analytic in terms of the vielbeins of the first order formulation of 3D gravity. As a result this term is not analytic in the Chern-Simons gauge fields. This could be easily resolved by setting $\b = 0$ in the generalized action. Before turning to the Chern-Simons theory and its reduction to a string with AdS$_3$ target spacetime we explore a more interesting alternative.


\subsection{Worldsheet metric} \label{suse:worldsheetmetric}

The area term in the generalized gravitational action~\eqref{se2:generalizedgraction} is not analytic in the Chern-Simons fields. In order to render the Chern-Simons formulation of the generalized action analytic in the gauge fields, let us introduce a boundary worldsheet metric $\g_{\mu\nu}$ so that the action becomes
  \eq{
  S_{GR} = \frac{1}{\k} \left \{ \frac{1}{2} \sint \absq{g} \(R + 2\) + \a \dsint \absq{h} K \pm \frac{\b}{2} \dsint \absq{\g} \g^{\mu\nu} h_{\mu\nu} \right \}. \label{se2:polyakovgraction}
  }
The worldsheet metric is a dynamical object, i.e.~not a background metric, whose equation of motion yields the following constraint at the \emph{asymptotic} boundary
  \eq{
  h_{\mu\nu} - \frac{1}{2} \g_{\mu\nu} \g^{\rho\s} h_{\rho\s} = 0. \label{se2:constraint}
  }
This equation can be used to restore the action~\eqref{se2:polyakovgraction} to its original form~\eqref{se2:generalizedgraction} and is similar to what is done in string theory to turn the Polyakov action into the Nambu-Goto action. The analogy is a close one since in the reduction of the Chern-Simons theory to the WZW/string theory, the kinetic term of the latter comes from the area term. It is for this reason that we must be careful with signs when taking square roots. The $\pm 1$ factor in the last term of eq.~\eqref{se2:polyakovgraction} originates from
  \eq{
  \absq{h} = \pm\frac{1}{2} \absq{\g} \g^{\rho\s} h_{\rho\s}.
  }

Since we have been careful about the validity of our actions, we should check if eq.~\eqref{se2:polyakovgraction} has a well-defined variational principle and a finite Brown-York stress tensor. Variation of eq.~\eqref{se2:polyakovgraction} yields,
  \begin{equation}
  \begin{split}
  \d S_{GR} &= \frac{1}{2\k} \sint \absq{g}\, \Pi_{\mu\nu} \d g^{\mu\nu} -\frac{1}{2} \dsint\absq{ h^{(0)}}\, \widetilde{T}^{BY}_{\mu\nu} \d h^{(0)\mu\nu} + \frac{\b}{2\k} \dsint \absq{\g}\, \pi_{\mu\nu} \d\g^{\mu\nu} \\
  &+ \frac{\b}{2\k} \dsint \( \absq{h}\, h_{\mu\nu} \mp \absq{\g}\, \g^{\rho\s} h_{\rho\mu} h_{\s\nu} \) \d h^{\mu\nu} - \frac{ 2 \a + \b - 1}{2\k} \dsint \absq{h^{(0)}} [ \dots ]
  \end{split} \label{se2:polyakovvariation}
  \end{equation}
where $\widetilde{T}^{BY}_{\mu\nu}$ is the Brown-York tensor given in eq.~\eqref{se2:bytmunuonshell}, $\pi_{\mu\nu} = h_{\mu\nu} - \frac{1}{2}\g_{\mu\nu} \g^{\rho\s} h_{\rho\s}$ is the constraint equation~\eqref{se2:constraint}, and $[\dots]$ is the term in brackets in the second line of  eq.~\eqref{se2:variationfinal}. The second line of eq.~\eqref{se2:polyakovvariation} must vanish if the action is to have a well-posed variational principle. The first term in the second line vanishes if the constraint holds while the second term vanishes if $2\a + \b =1$. Thus, the action~\eqref{se2:polyakovgraction} is well-defined provided that $2\a + \b =1$ and the bulk equations of motion $\Pi_{\mu\nu} = 0$ \emph{and} constraints $\pi_{\mu\nu} = 0$ hold. 

Unlike the analysis of the previous section, where both Brown-Henneaux and free boundary conditions were applicable, the action~\eqref{se2:polyakovgraction} is compatible only with the latter. In order to see this let us take a closer look at the constraint equation. Using the parametrization of the metric given in eqs.~\eqref{se2:bcgauge} and~\eqref{se2:inducedmetric} -- which is valid for both free and Brown-Henneaux boundary conditions -- we find that
  \eq{
  h^{(j)}_{\mu\nu} - \frac{1}{2} \g_{\mu\nu} \g^{\rho\s} h^{(j)}_{\rho\s} = 0, \hspace{30pt} j = 0, 2. \label{se2:constraint2}
  }
The separate constraints for leading and subleading components are imposed because eq.~\eqref{se2:constraint} holds at the asymptotic boundary and the $h^{(j)}_{\mu\nu}$ terms are independent of the radial coordinate, cf.~eq.~\eqref{se2:inducedmetric}. Since the constraint holds asymptotically subleading components with $j > 2$ are not constrained.\footnote{This is a consequence of the special status of the asymptotic boundary in the metric formulation of 3D gravity. In contrast, in the Chern-Simons theory it is possible to have boundaries at finite values of the radial coordinate. We only study the case where both formulations are meaningful.}

For the leading component of the induced metric we can always choose $h^{(0)}_{+-} = e^{2\vp} \g_{+-}$ since the constraint equation~\eqref{se2:constraint2} leaves $h^{(j)}_{+-}$ undetermined. Here $\pm$ denote indices of lightcone coordinates, i.e.~$x^{\pm} = t \pm \phi$. The constraint equation for the leading components then becomes
  \eq{
  h^{(0)}_{\mu\nu} = e^{2\vp} \g_{\mu\nu}. \label{se2:constraintleading}
  }
Thus, choosing a gauge for the worldsheet metric  is equivalent to fixing a gauge for the boundary metric up to a conformal factor. We can only do this for $h^{(0)}_{\mu\nu}$ since the bulk equations of motion determine the subleading components of the metric in terms of its leading components. This is most easily seen in the Fefferman-Graham gauge~\cite{Fefferman:1985ok} where the full metric reads
  \eq{
  ds^2 = \frac{1}{r^2} dr^2 + h_{\mu\nu} dx^{\mu} dx^{\nu}, \label{se2:fggauge}
  }
and $h_{\mu\nu}$ is the induced metric given in eq.~\eqref{se2:inducedmetric}. The bulk equations of motion yield (see for example~ref.~\cite{Skenderis:1999nb}),
  \eq{
  h^{(2)}_{\mu\nu} = -\frac{1}{2} \( h^{(0)}_{\mu\nu} R^{(0)} - T_{\mu\nu} \), \hspace{30pt} h^{(4)}_{\mu\nu} = \frac{1}{4} h^{(2)}_{\mu\rho} h^{(0)\rho\s} h^{(2)}_{\s\nu}, \label{se2:fggauge2}
  }
while all other subleading components vanish. The rank-two tensor $T_{\mu\nu}$ is proportional to the covariant expression for the Brown-York tensor given in eq.~\eqref{se2:bytmunu} and satisfies,
  \eq{
  \nabla^{(0)}_{\mu} T^{\mu}_{\phantom{\mu}\nu} = 0, \hspace{30pt} h^{(0)\mu\nu} T_{\mu\nu} = R^{(0)}. \label{se2:fggauge3}
  }
Thus, once we fix the gauge for the boundary metric $h^{(0)}_{\mu\nu}$ the equations of motion determine the full metric up to two \emph{integration constants} that depend on the boundary coordinates. 

The consequences of the constraint equation on the subleading $h^{(2)}_{\mu\nu}$ components of the metric are more dramatic. The reason is that the subleading components determine the charges of the solutions of the theory. Using eq.~\eqref{se2:constraintleading} the constraint equation for the subleading components of the metric becomes, provided that $h^{(0)}_{+-} \ne 0$,
  \eq{
  0 = h^{(2)}_{\mu\nu} - h^{(0)}_{\mu\nu}\, \frac{h^{(2)}_{+-}}{h^{(0)}_{+-}}, \label{se2:constraintsubleading}
  }
where we note that $h^{(2)}_{+-}$ is left undetermined. This equation is compatible with the equations of motion that follow from a dynamical boundary metric, namely with $\widetilde{T}^{BY}_{\mu\nu} = 0$.\footnote{When the worldsheet is flat it is possible to reproduce $T^{BY}_{\mu\nu} = 0$ for all components of the Brown-York tensor as we will see in Section \ref{se:cstowzw}.} Indeed, using eq.~\eqref{se2:bytmunuonshell} we find
  \eq{
  0 = \widetilde{T}^{BY}_{\mu\nu} = \frac{1}{\k} \( h^{(2)}_{\mu\nu} - h^{(0)}_{\mu\nu} \,\frac{h^{(2)}_{+-}}{h^{(0)}_{+-}} \), \label{se2:bytmunu2}
  }
where we have \emph{integrated in} the $\widetilde{T}^{BY}_{+-} = 0$ component of the equations of motion into $\widetilde{T}^{BY}_{\pm\pm} = 0$. Thus, we conclude that replacing the square-root structure of the area term with the worldsheet metric is consistent only if the boundary metric $h^{(0)}_{\mu\nu}$ is dynamical, i.e.~if free boundary conditions hold. This conclusion follows directly from the constraint equation~\eqref{se2:constraint}, which was necessary to make the generalized gravitational action~\eqref{se2:polyakovgraction} well-defined. Note that the constraint equation cannot reproduce the $\widetilde{T}^{BY}_{+-} = 0$ component of the equations of motion because the last term in eq.~\eqref{se2:polyakovgraction} is Weyl invariant.


\section{Asymptotic dynamics} 
\label{se:cstowzw}

In this section we show that the asymptotic dynamics of AdS$_3$ gravity with free boundary conditions are described by a string with target spacetime AdS$_3$. We begin by considering the Chern-Simons formulation of the generalized gravitational action and then turn to its reduction to the SL(2,R) WZW/string theory. We show that constraints on the SL(2,R) currents do not implement boundary conditions on the metric, in contrast with previous approaches. We also check that the space of metric solutions satisfying the bulk equations of motion and the Virasoro constraints is non-trivial.

\subsection{Chern-Simons formulation} \label{suse:cs}

Let us now consider the Chern-Simons formulation of the generalized gravitational action given in eq.~\eqref{se2:polyakovgraction}. As shown in refs.~\cite{Achucarro:1987vz,Witten:1988hc} an alternative description of three dimensional gravity with a negative cosmological constant is given by two Chern-Simons actions with opposite sign. Keeping track of boundary terms we have,
  \eq{
  \sint \absq{g}\( R + 2 \) = I_{CS}[A] - I_{CS}[\bar{A}] - \dsint \tr \big ( A \we \bar{A} \big ), \label{se3:ehcs}
  }
where $A$ and $\bar{A}$ are two SL(2,R) gauge fields defined in terms of the vielbein $e^a$ and dual spin connection $\om^a = \frac{1}{2} \e^{abc} \om_{bc}$ by~\cite{Witten:1988hc} \footnote{Our conventions for the generators $T^a$ of SL(2,R) are the following:  $[T^a,T^b] = \e^{abc} \eta_{cd} T^d$ where $\e^{123} = 1$, and $\mathrm{tr} \( T^a T^b \) = \frac{1}{2} \eta^{ab}$ where $\eta_{ab} = \mathrm{diag}\,\(-,+,+\)$. We also assume that the determinant of the vielbein is positive, i.e.~$\absq{g} = \mathrm{det}(e^{a}_{\phantom{a}\mu})$; this determines the overall sign in front of the Einstein-Hilbert action.}
  \eq{
  A^a = \( w^a + e^a \), \hspace{30pt} \bar{A}^a  = \( w^a - e^a \), \label{se3:vielbeins}
  }
and $I_{CS}$ is the Chern-Simons action given by
  \eq{
  I_{CS}[A] = \sint \tr \Big (A\we d A + \frac{2}{3} A\we A \we A \Big ).
  }

The boundary term in eq.~\eqref{se3:ehcs} is proportional to the Gibbons-Hawking term, namely $\dsint \tr (A\we \bar{A}) = \dsint e^a \we \om^a = \dsint \absq{h} K$. Thus, the Chern-Simons formulation of the generalized gravitational action~\eqref{se2:polyakovgraction} is given by
  %
  \eq{
  S_{CS} = \frac{k}{4\pi} \left \{ I_{CS}[A] - I_{CS}[\bar{A}] + J_{\a,\b}[A,\bar{A}] \right\}, \label{se3:csaction}
  }
where $k = 1/4G_N$ and the boundary term $J_{\a,\b}[A,\bar{A}]$ reads
  \eq{
  J_{\a,\b}[A,\bar{A}] = \( 2\a - 1 \) \dsint \tr (A\we \bar{A}) \pm \frac{\b}{2} \dsint \absq{\g} \g^{\mu\nu}\tr \big [ \( A - \bar{A}\)_{\mu}\(A - \bar{A}\)_{\nu}\big ]. \label{se3:csboundary}
  }

Note that all boundary terms in eq.~\eqref{se3:csboundary} are covariant and analytic in the gauge fields. In particular the worldsheet metric acts as the boundary metric in the Chern-Simons theory. As shown in the previous section, fixing a gauge for the worldsheet metric translates through the constraint equations to fixing a gauge (up to a conformal factor) for the boundary metric $h^{(0)}_{\mu\nu}$ of the metric theory~\eqref{se2:constraintleading}. Thus, both the metric theory and the Chern-Simons theory, whose area term (the last term in eq.~\eqref{se3:csboundary}) is Weyl invariant, see the same conformal structure at the boundary.\footnote{However note that the relative conformal factor will show up in the conformal anomaly.}

Now, while it was not strictly necessary to introduce a worldsheet metric into the action, i.e.~we could have set $\b = 0$ in eq.~\eqref{se2:generalizedgraction}, the area term above is crucial in the reduction of the Chern-Simons theory to that of a string propagating in AdS$_3$. This was originally noted by Arcioni, Blau, and O'Loughlin~\cite{Arcioni:2002vv} who considered the action~\eqref{se3:csaction} with $\a =0$, $\b = 1$, and $\g_{\mu\nu}$ a non-dynamical, flat metric. There, it was argued that the action of Chern-Simons formulation of 3D gravity should be invariant under diagonal SL(2,R) $\times$ SL(2,R) transformations of the gauge fields. This is a reasonable requirement since these transformations correspond to local Lorentz transformations of the vielbeins, \emph{gauge} symmetries of the first order formulation of gravity. It is only for $\a = 0$ that local Lorentz invariance is a symmetry of the action~\eqref{se3:csaction}.

We will therefore fix $\a = 0$. A well-defined action in the metric formalism is recovered if $\b = 1 - 2\a = 1$. We can always do this since, as shown in the previous section, the boundary terms accompanying the Einstein-Hilbert action are not unique. Finally, we choose the lower (minus) sign in the last term of eq.~\eqref{se3:csboundary} since it leads to a healthy kinetic term in the WZW theory. Thus, the action we will consider is given by
  \eq{
  S_{GR} = \frac{k}{4\pi} \left \{ I_{CS}[A] - I_{CS}[\bar{A}] - \dsint A\we \bar{A} - \frac{1}{2} \dsint \absq{\g} \g^{\mu\nu} \( A - \bar{A}\)_{\mu}\(A - \bar{A}\)_{\nu} \right \}. \label{se3:finalcsaction}
  }
The new element in the action~\eqref{se3:finalcsaction} compared to that of~\cite{Arcioni:2002vv} is the dynamical worldsheet metric $\g_{\mu\nu}$ which we have found to be compatible only with free boundary conditions. Note that we could have introduced the worldsheet metric directly in the Chern-Simons formulation of~\cite{Arcioni:2002vv} by demanding not only local Lorentz invariance of the vielbein, but also diffeomorphism invariance of the boundary.


\subsection{Reduction to WZW/string theory} \label{suse:wzw}

Another motivation behind the choice $\a = 0$, $\b = 1$ is the reduction to the WZW/string theory to which we now turn. The equations of motion of the Chern-Simons theory, namely,
  \eq{
  0 = F = d A + A\we A, \hspace{30pt} 0 = \bar{F} = d \bar{A} + \bar{A} \we \bar{A},
  }
imply that all solutions are locally of the form,
  \eq{
  A = g^{-1} d g, \hspace{30pt} \bar{A} = \bar{g}^{-1} d \bar{g}, \label{se3:puregauge}
  }
where $g$ and $\bar{g}$ are elements of SL(2,R). Not all of these solutions are pure gauge, however, since the presence of a boundary \emph{spontaneously breaks} the gauge symmetries -- some of the hitherto pure gauge modes become physical. To be more precise, some of the generators of gauge transformations become generators of global symmetries that map physical states into physical states (see ref.~\cite{Banados:1998gg} for a pedagogical discussion). This statement is sensitive to boundary conditions since the latter determine which of the generators $g$ and $\bar{g}$ of SL(2,R) $\times$ SL(2,R) become generators of global symmetries. The corresponding modes given by eq.~\eqref{se3:puregauge} become physical, i.e.~they have finite, non-vanishing charges.

A similar story unfolds in the metric theory whose linearized perturbations are pure gauge $\d g_{MN} = \nabla_{(M} \xi_{N )}$ for generic generators of diffeomorphisms $\xi^{M}$. This is the statement that 3D gravity, like Chern-Simons theory, has no local degrees of freedom. However, a careful choice of boundary conditions renders some of the gauge symmetries generated by $\xi^{M}$ true global (but asymptotic) symmetries. The linearized perturbations $\d g_{MN}$ corresponding to these vectors are physical, i.e.~they have finite charges. Like the physical solutions to the Chern-Simons theory, the linearized perturbations of gravity in AdS$_3$ only interact at the boundary since their coupling to sources is a total derivative.

Therefore, one expects the dynamics of 3D gravity to be captured by a theory at the boundary. This is most easily seen in the Chern-Simons formulation by integrating in the equations of motion $F = \bar{F} = 0$, i.e.~by substituting eq.~\eqref{se3:puregauge} into the action~\eqref{se3:finalcsaction}~\cite{Moore:1989yh,Elitzur:1989nr}.\footnote{This is a covariant generalization of the procedure used in refs.~\cite{Moore:1989yh,Elitzur:1989nr}. The reason we are able to integrate in the equations of motion in the Chern-Simons formulation of the theory is that one of the components of the Chern-Simons gauge field acts as a Lagrange multiplier. Indeed, if we write one of the Chern-Simons actions as $I_{CS} = \sint  \( 2 A_r F_{\phi t}  + A_t \p_r A_{\phi} - A_{\phi} \p_r A_t \) $ we see that $A_r$ enters as a Lagrange multiplier enforcing the $F_{\phi t} = 0$ constraint in the path integral. Note that both this form of the Chern-Simons action and the covariant one given in eq.~\eqref{se3:finalcsaction} lead to the same results since we drop total derivatives with respect to the $t$ and $\phi$ coordinates.} We then find $S_{GR}[A,\bar{A}] \ra S_{WZW}[g,\bar{g}]$ where
  \eq{
  S_{WZW} = W[g] + W[\bar{g}^{-1}] -  \frac{k}{4\pi} \dsint \gg^{\mu\nu}_{-} \( g^{-1} \p_{\mu} g\) \( \bar{g}^{-1} \p_{\nu} \bar{g} \). \label{se3:prewzw}
  }
Here $W[g]$ stands for the WZW action
  \eq{
  W[g] = -\frac{k}{4\pi} \left \{   \frac{1}{2} \dsint \absq{\g} \g^{\mu\nu} \( g^{-1} \p_{\mu} g\) \( g^{-1} \p_{\nu} g\) + \frac{1}{3} \sint \( g^{-1} d g \)^3\right  \}, \label{se3:wzw}
  }
and $\gg^{\mu\nu}_{-}$ is a mixed tensor defined by 
  \eq{
  \gg^{\mu\nu}_{\pm} = \e^{\mu\nu} \pm \absq{\g} \g^{\mu\nu}.
  }
The right-hand side of eq.~\eqref{se3:prewzw} is a generalization of the Polyakov-Wiegman identities~\cite{Polyakov:1984et} for a WZW action with arbitrary metric $\g_{\mu\nu}$, namely
  \eq{
  W[g\bar{g}^{-1}] = W[g] + W[\bar{g}^{-1}] -  \frac{k}{4\pi} \dsint \gg^{\mu\nu}_{-} \( g^{-1} \p_{\mu} g\) \( \bar{g}^{-1} \p_{\nu} \bar{g} \). 
  }
  %
  


The asymptotic dynamics of three-dimensional gravity with free boundary conditions are thus described by
  \eq{
  S_{WZW} = -\frac{k}{4\pi} \left \{ \frac{1}{2} \dsint \absq{\g} \g^{\mu\nu} \tr \big [ (G^{-1} \p_{\mu} G )(G^{-1} \p_{\nu} G ) \big ] + \frac{1}{3} \int_{\ss} \tr \(G^{-1} d G \)^3 \right \}, \label{se3:string}
  }
where $\g_{\mu\nu}$ is a dynamical metric and $G \in$ SL(2,R) is given by
  \eq{
  G = g \bar{g}^{-1}. \label{se3:wzwfield}
  }
This is the action of a string whose target spacetime is the group manifold of SL(2,R), namely AdS$_3$.\footnote{At least locally since we have used $A = g^{-1} d g$ and $\bar{A} = \bar{g}^{-1} d \bar{g}$ in the reduction to the WZW theory.} The worldsheet coordinates of this string correspond to the boundary coordinates of the bulk spacetime where the Chern-Simons theory is defined. Similar observations were made in~\cite{Arcioni:2002vv} although their construction lacked the Virasoro constraints of string theory. In particular, the observation that for $\a=0$ and $\b = \pm 1$ the WZW action can be obtained directly from the Chern-Simons action by means of the Polyakov-Wiegmann identities was presented there. Since our worldsheet metric is a dynamical object we obtain the constraints
  \eq{
  0 = -\frac {2}{\absq{\g} }\frac{\d S_{WZW}}{\d \g^{\mu\nu}} = T^{S}_{\mu\nu}, \label{se3:virconstraints}
  }
where $T^{S}_{\mu\nu}$ is the Sugawara stress-energy tensor of the WZW model. 

Note that the factorization of the WZW group element $G$ in eq.~\eqref{se3:wzwfield} is inherited from the Chern-Simons theory and constitutes an extra layer of interpretation on WZW solutions. The fact that Chern-Simons boundary conditions put constraints on $g$ and $\bar g$, which lack direct meaning in the WZW model, will lead to the compatibility conditions studied in the next section.


\subsection{Equations of motion, SL(2,R) currents, and Virasoro constraints} \label{suse:onshell}

We have seen that an SL(2,R) WZW model with Virasoro constraints captures the asymptotic dynamics of three-dimensional gravity with free boundary conditions. The equations of motion of the WZW model impose constraints on the group elements $g$ and $\bar{g}$ parametrizing the WZW and Chern-Simons variables. This is because solutions to the WZW model are not generically compatible with the boundary conditions obeyed by the Chern-Simons fields. In order to see this let us compute the on-shell values of the WZW and Chern-Simons fields. These will be used next to compute the SL(2,R) currents and Virasoro constraints of the WZW model. 

We begin by gauge fixing the worldsheet metric to a flat metric, which is always possible at the classical level,
  \eq{
  \g_{\mu\nu} = \eta_{\mu\nu}, \hspace{30pt} \eta_{+-} = -\frac{1}{2}, \quad \eta_{\pm\pm} = 0. \label{se3:gaugefixing}
  }
Variation of the Chern-Simons action~\eqref{se3:finalcsaction} then yields,
  \eq{
  \d S_{CS} = \frac{k}{2\pi} \left \{  \sint \tr \( \d A \we F + \d \bar{A} \we \bar{F} \) - 2 \dsint \tr \left ( e_- \d A_+ - e_+ \d \bar{A}_- \right ) \right \},
  }
where $e_{\mu} = \frac{1}{2} (A - \bar{A} )_{\mu}$. Since the $e_{\pm}$ components of the vielbeins determine the induced metric via
  \eq{
  h_{\mu\nu} = 2\, \tr \( e_{\mu} e_{\nu} \) = \frac{1}{2} \tr \left [  (A - \bar{A} )_{\mu} (A - \bar{A} )_{\nu} \right], \label{se3:metric}
  }
we do not want to set them to zero. A well-posed variational principle is recovered provided that $A_+$ and $\bar{A}_-$ are held fixed at the boundary,
  \eq{
  \d A_{+} = 0, \hspace{30pt} \d \bar{A}_{-} = 0. \label{se3:csbc}
  }
These boundary conditions constrain the possible values of $g$ and $\bar{g}$ in the parametrization of the Chern-Simons gauge fields used in eq.~\eqref{se3:puregauge}. They also constrain the asymptotic values of the parameters $u$, $\bar{u}$ of SL(2,R) gauge transformations. Indeed, under a gauge transformation,
  \eq{
  A_+ \ra A^{u}_+ = (gu)^{-1} \p_+ (gu), \hspace{30pt} \bar{A}_- \ra \bar{A}^{\bar{u}}_- = (\bar{g} \bar{u})^{-1} \p_- (\bar{g} \bar{u}).
  }
This implies that, asymptotically,
  \eq{
  g &= g_-(x^-,r) v(x^{\pm},r), \hspace{-40pt}  & u  = v(x^{\pm},r)^{-1} u_-(x^-,r), \label{se3:bc1}\\
  \bar{g} &= \bar{g}_+(x^+,r) \bar{v}(x^{\pm},r), \hspace{-40pt} & \bar{u} = \bar{v}(x^{\pm},r)^{-1} \bar{u}_+(x^+,r). \label{se3:bc2}
  }

On the other hand the equations of motion of the WZW model are given by
  \eq{
  \p_- \( G^{-1} \p_+G \) = 0.
  }
The solutions to this equation are not generically compatible with both the parametrization $G = g \bar{g}^{-1}$ used in the reduction of the Chern-Simons theory and the boundary conditions obeyed by the Chern-Simons fields. Indeed, $G = g \bar{g}^{-1}$ is not a solution to the WZW equations of motion unless $v$ and $\bar{v}$ in eqs.~\eqref{se3:bc1} and~\eqref{se3:bc2} are independent of $x^+$ and $x^-$, respectively. Thus, compatibility with the boundary conditions in the Chern-Simons theory yields the following on-shell parametrization of the WZW and Chern-Simons variables,
  \eq{
  G = g_{-}^{\phantom{+}} \bar{g}^{-1}_+, \hspace{30pt} A = g_-^{-1} d g_-^{\phantom{1}},  \hspace{30pt}  \bar{A} = \bar{g}_+^{-1} d \bar{g}_+^{\phantom{1}}, \label{se3:onshell}
  }
where we have removed the now redundant $v(x^-,r)$ and $\bar{v}(x^+,r)$ functions by shifts in $g_-$ and $\bar{g}_+$, respectively.
In particular note that the $A_+$ and $\bar{A}_-$ components of the gauge fields vanish.

\subsection*{SL(2,R) currents} \label{suse:sl2rcurrents}

We can now compute the on-shell values of the SL(2,R) currents of the WZW model~\eqref{se3:string}. Up to a normalization the SL(2,R) currents are given by,
  \eq{
  J_+ &= k \,G^{-1} \p_+ G = k\,\bar{g}^{\phantom{1}}_+ \p_+ \bar{g}^{-1}_+,  \hspace{-30pt}  &J_- = k\, G \p_- G^{-1} = k\, g_{-}^{\phantom{1}} \p_- g^{-1}_-.
  }
Note that this is not the same identification between the group elements $g$, $\bar{g}$ and the gauge fields $A$, $\bar{A}$ given in eq.~\eqref{se3:onshell}. In other words, the SL(2,R) currents are not proportional to the Chern-Simons gauge fields. Boundary conditions on the latter, which are equivalent to imposing boundary conditions on the vielbeins and spin connection, determine boundary conditions on the metric of 3D gravity. Thus, we are led to conclude that any constraints on the currents of the WZW model would not directly enforce boundary conditions on the metric. We will see momentarily how this contrasts with previous approaches.

To illustrate this consider $g = g_- (x^-,r)$ and $\bar{g} = \bar{g}_+(x^+,r)$ so that the boundary conditions~\eqref{se3:csbc} on the Chern-Simons fields are satisfied. Suppose these also lead to a metric satisfying free boundary conditions~\eqref{se2:freebc}. Note that $G = g_-\bar{g}_+^{-1}$ solves the equations of motion of the WZW model. From this solution it is possible to construct other solutions by letting,
  \eq{
 g_- \ra V g_-,  \hspace{30pt}  \bar{g}_+ \ra  U \bar{g}_+, \label{se3:shiftedsol}
  }
where $U$ and $V$ are constant elements of SL(2,R). Clearly, the new solutions leave the Chern-Simons fields invariant. Therefore they do not change the boundary conditions of the corresponding metric. In fact, they do not change the metric at all. The new solutions \emph{do} change the SL(2,R) currents, however,
  \eq{
  J_+ = k\, U \bar{g}^{\phantom{1}}_+ \p_+ \bar{g}^{-1}_+ U^{-1},  \hspace{30pt}  J_- = k\, V g_{-}^{\phantom{1}} \p_- g^{-1}_- V^{-1}.
  }
In particular, one can choose a basis such that, for some generator $T^a$, the zero modes of $J_+^a \pm J_-^a$ correspond to the energy and angular momentum of the string in the string target spacetime~\cite{Maldacena:2000hw}. Thus the solutions~\eqref{se3:shiftedsol} are physically distinct in the string theory. 

On the other hand local Lorentz transformations of the vielbeins, i.e.~transformations for which
  \eq{
  g_- \ra g_- U,  \hspace{30pt}  \bar{g}_+ \ra \bar{g}_+ U,
  }
where $U$ is a constant element of SL(2,R) do change the Chern-Simons gauge fields but do not change the SL(2,R) currents. This is to be expected since this is a redundancy in the first order formulation of general relativity. Accordingly, these transformations are trivial in the WZW model.
  
In the standard construction of Coussaert, Henneaux, and van Driel~\cite{Coussaert:1995zp} the currents of the WZW model are proportional to the fundamental gravitational fields of the first order formulation of gravity. Thus, constraints on the former enforce the Brown-Henneaux boundary conditions~\cite{Brown:1986nw} on the latter. In contrast, we have found that free boundary conditions are not implemented through constraints linear on the SL(2,R) currents. 
This means that the WZW theory is not automatically reduced to Liouville theory. 
However, the Virasoro constraints~\eqref{se3:virconstraints} render the spectrum of the theory unitary~\cite{Hwang:1990aq,Evans:1998qu,Maldacena:2000hw}.\footnote{Note, however, that the string theory is not critical unless new degrees of freedom are added to the theory.}

\subsection*{Virasoro constraints} \label{suse:virconstraints}

Although the SL(2,R) currents do not directly implement boundary conditions on the metric, they still impose constraints on the latter through the Virasoro constraints. Indeed, for a flat worldsheet metric eq.~\eqref{se3:virconstraints} implies, in the semiclassical, large-$k$ limit, 
  \eq{
  0 &= T^{S}_{++} = \frac{1}{k} \tr(J_{+} J_{+}) = k\, \tr (\bar{A}_{+} \bar{A}_{+}) = 2 k \,h_{++},  \label{se3:constraint1} \\
    0 &= T^{S}_{--} = \frac{1}{k} \tr(J_{-} J_{-}) = k\, \tr (A_{-} A_{-}) = 2 k \,h_{--},  \label{se3:constraint2}
  }
where we have used the on-shell values of $G$, $A$ and $\bar{A}$ given in eq.~\eqref{se3:onshell}. These are precisely the constraints studied in Section~\ref{suse:worldsheetmetric} when the worldsheet metric is flat, cf.~eq.~\eqref{se2:constraint}. As discussed there, eqs.~\eqref{se3:constraint1} and~\eqref{se3:constraint2} constrain the leading and first subleading components of the induced metric. For the leading components we have~\eqref{se2:constraintleading},
  \eq{
  h^{(0)}_{\mu\nu} = e^{2\vp} \eta_{\mu\nu}, \label{se3:leadingconstraint}
  }
that is, fixing a gauge for the worldsheet metric fixes a gauge for the boundary metric up to a conformal factor, while for the subleading components we find~\eqref{se2:constraintsubleading},
  \eq{
  h^{(2)}_{\pm\pm} = 0, \label{se3:subleadingconstraint}
  }
which is consistent with vanishing of the Brown-York tensor~\eqref{se2:bytmunu2}.

One may think that satisfying the two constraints~\eqref{se3:subleadingconstraint} removes all but one solution from the theory. The reason being that the most general solution to Einstein's equations can be written down in closed form using only two arbitrary functions. Indeed, in the Fefferman-Graham gauge~\eqref{se2:fggauge} one finds, for \emph{Brown-Henneaux} boundary conditions, 
  \eq{
  ds^2 = \frac{dr^2}{r^2} - r^2\( dx^+ - \frac{1}{r^2} f_- dx^- \) \( dx^- - \frac{1}{r^2} f_+ dx^+  \), \label{se3:bhfg}
  }
where $f_{\pm} = f_{\pm}(x^{\pm})$ parametrize the space of solutions to the equations of motion. Then the constraint equation would imply that only the $f_{\pm} = 0$ solution survives. However, as seen from eq.~\eqref{se3:leadingconstraint} the Virasoro constraints fix the boundary metric only up to a conformal factor. In that case eq.~\eqref{se3:bhfg} is not the most general solution to the equations of motion. 

While eq.~\eqref{se3:leadingconstraint} is not compatible with Brown-Henneaux boundary conditions, it is consistent with free boundary conditions where the boundary metric is in the conformal gauge~\cite{Troessaert:2013fma}. In this case the most general solution to the equations of motion can be obtained from eqs.~\eqref{se2:fggauge} and~\eqref{se2:fggauge2} by solving eq.~\eqref{se2:fggauge3}. For the subleading components we have,
  \eq{
  h^{(2)}_{\pm\pm} = f_{\pm}(x^{\pm}) - \( \p_{\pm} \vp \p_{\pm} \vp - \p^2_{\pm} \vp \),  \hspace{30pt}  h^{(2)}_{+-} = \p_+\p_- \vp, \label{se3:phasespace}
  }
where the second term is proportional to the stress-energy tensor $T^{L}_{\pm\pm}$ of Liouville theory with a vanishing cosmological constant.  Thus, while the Virasoro constraints do require $h^{(2)}_{\pm\pm}$ to vanish, the space of solutions is not empty but constrained to obey $f_{\pm} = T^{L}_{\pm\pm}$. 

Now recall that the Virasoro constraints do not reproduce the vanishing of the $\widetilde{T}^{BY}_{+-}$ component of the Brown-York tensor, as required by free boundary conditions. The reason for this is that the area term in the Chern-Simons action~\eqref{se3:finalcsaction} and the WZW action~\eqref{se3:string} are Weyl invariant. However, one may show that eq.~\eqref{se3:onshell}, in particular the chirality of the group elements $g_-$ and $\bar{g}_+$, leads to $h^{(2)}_{+-} = 0$. The latter is proportional to the $\widetilde{T}^{BY}_{+-}$ component of the Brown-York stress tensor. Thus the theory is compatible with all of the equations of motion/constraints that result by imposing free boundary conditions.

Let us conclude this section by mentioning that, since we are dealing with a non-critical string whose central charge is $c = 3k/(k + 2) \sim 3$ in the semiclassical, large-$k$ limit, we do not expect to be able to fix the worldsheet metric as in eq.~\eqref{se3:gaugefixing} in the quantum theory. A consistent treatment of the theory may require additional degrees of freedom at the boundary, either extra \emph{dimensions} to the string theory, or a Liouville dimension originating from the conformal factor of the worldsheet metric. In this case eq.~\eqref{se3:phasespace} will receive further contributions so that $h^{(2)}_{\pm\pm}$ no longer vanishes, i.e.
  \eq{
  h^{(2)}_{\pm\pm} + T^{new}_{\pm\pm} = 0,
  }
where $T^{new}_{\mu\nu}$ is the stress-energy tensor of the additional degrees of freedom. This will lead to finite charges in the metric theory since these are proportional to the $h^{(2)}_{\pm\pm}$ components of the metric.


\section{Discussion and conclusions}
\label{se:conclusions}

In this paper we have shown that the asymptotic dynamics of pure AdS$_3$ gravity with free boundary conditions are described by a string propagating on an AdS$_3$ target spacetime. One of the challenges in the derivation of this result was finding the appropriate covariant boundary terms in the Chern-Simons formulation of 3D gravity. The reduction of the resulting Chern-Simons theory~\eqref{se3:finalcsaction} led to an SL(2,R) WZW model with Virasoro constraints, i.e.~to the action of a string in AdS$_3$. We have seen that the on-shell values of the WZW currents do not implement boundary conditions on the theory. Nevertheless, the Virasoro constraints of the WZW model do enforce boundary conditions to some extent, i.e.~they fix the boundary metric up to a conformal factor. More dramatically, the Virasoro constraints also imply vanishing of the subleading components of the metric. We have shown that this does not lead to an empty spectrum, however, and the theory still describes non-trivial metric solutions.


We note that the string describing the asymptotic dynamics of 3D gravity with free boundary conditions is non-critical, i.e.~its central charge is $c = 3$ in the semiclassical limit. Therefore the conformal anomaly will play a role in the quantization of the theory. The corresponding classical treatment requires adding new dimensions to the string theory, i.e.~new degrees of freedom at the boundary, or dealing with the Liouville dimension that arises from the conformal factor of the worldsheet metric. In either case the contributions to the Virasoro constraints from these new degrees of freedom allow for non-vanishing subleading components of the metric.

The connection between pure AdS$_3$ gravity and string theory on AdS$_3$ has the potential to help us understand aspects of the spectrum of pure 3D gravity. It is then necessary to continue the group elements $g_-$ and $\bar{g}_+$ defined asymptotically in eqs.~\eqref{se3:bc1} and~\eqref{se3:bc2} to finite values of the bulk radial coordinate. This continuation is not determined by the equations of motion of the WZW model. One way to do this is to follow the prescription of~\cite{Coussaert:1995zp} but other, more general prescriptions may be possible. 

Once this is figured out we do not expect all solutions to the WZW/string theory to correspond to appropriate metrics. Some of the solutions may violate the boundary conditions of the metric, e.g.~on the $g_{r\mu}$ components, in which case the latter must be imposed as constraints on the spectrum of the theory. The problem of interpreting solutions in the WZW model is an old one and it already arises at the level of the Chern-Simons theory~\cite{Witten:2007kt}. Whether or not solutions with unhealthy, e.g.~degenerate, metrics should be considered as part of the spectrum of 3D gravity will depend on what can be learned from them. 

Note also that the symmetries of the WZW model are generically larger than those of the metric theory. Indeed, when the worldsheet metric is flat, the WZW model has left and right-moving $\widehat{\mathrm{SL}}$(2,R) symmetries. On the other hand the symmetries in the metric formalism consist of left and right-moving $\widehat{\mathrm{U}}$(1) $\times$ Virasoro symmetries~\cite{Troessaert:2013fma}. From this point of view a restriction on the spectrum of the WZW/string theory may be necessary as well.

Clearly, many consequences of our approach to 3D gravity are still uncertain, but to conclude we have established a number of basic facts. First, a string with AdS$_3$ target geometry can be used to describe pure 3D gravity with a negative cosmological constant. Second, this description translates asymptotic statements between 3D geometry and the 2D boundary in a different way than previous approaches. Third, the string description corresponds to free boundary conditions which generalize the standard Brown-Henneaux boundary conditions. It is therefore a subtly different physical model, which can have new physical consequences that we plan to explore in the future.

\section*{Acknowledgments}

BS wishes to thank Igor Klebanov and Juan Maldacena for comments at an early stage of this work.

\bibliographystyle{utphys2}
\bibliography{3dgravity}

\providecommand{\href}[2]{#2}\begingroup\raggedright\begin{thebibliography}{10}

\bibitem{Deser:1983tn}
S.~Deser, R.~Jackiw, and G.~'t~Hooft, ``{Three-Dimensional Einstein Gravity:
  Dynamics of Flat Space},''
{\em Annals Phys.} {\bfseries 152} (1984) 220.

\bibitem{Deser:1983nh}
S.~Deser and R.~Jackiw, ``{Three-Dimensional Cosmological Gravity: Dynamics of
  Constant Curvature},''
{\em Annals Phys.} {\bfseries 153} (1984) 405--416.

\bibitem{Brown:1986nw}
J.~D. Brown and M.~Henneaux, ``{Central Charges in the Canonical Realization of
  Asymptotic Symmetries: An Example from Three-Dimensional Gravity},''
{\em Commun.Math.Phys.} {\bfseries 104} (1986) 207--226.

\bibitem{Banados:1992wn}
M.~Ba\~nados, C.~Teitelboim, and J.~Zanelli, ``{The Black hole in
  three-dimensional space-time},'' {\em Phys.Rev.Lett.} {\bfseries 69} (1992)
  1849--1851,
\href{http://arxiv.org/abs/hep-th/9204099}{{\ttfamily arXiv:hep-th/9204099
  [hep-th]}}.

\bibitem{Banados:1992gq}
M.~Ba\~nados, M.~Henneaux, C.~Teitelboim, and J.~Zanelli, ``{Geometry of the
  (2+1) black hole},'' {\em Phys.Rev.} {\bfseries D48} no.~6, (1993)
  1506--1525,
\href{http://arxiv.org/abs/gr-qc/9302012}{{\ttfamily arXiv:gr-qc/9302012
  [gr-qc]}}.

\bibitem{Maldacena:1997re}
J.~M. Maldacena, ``{The Large N limit of superconformal field theories and
  supergravity},'' {\em Int.J.Theor.Phys.} {\bfseries 38} (1999) 1113--1133,
\href{http://arxiv.org/abs/hep-th/9711200}{{\ttfamily arXiv:hep-th/9711200
  [hep-th]}}.

\bibitem{Gubser:1998bc}
S.~Gubser, I.~R. Klebanov, and A.~M. Polyakov, ``{Gauge theory correlators from
  noncritical string theory},'' {\em Phys.Lett.} {\bfseries B428} (1998)
  105--114,
\href{http://arxiv.org/abs/hep-th/9802109}{{\ttfamily arXiv:hep-th/9802109
  [hep-th]}}.

\bibitem{Witten:1998qj}
E.~Witten, ``{Anti-de Sitter space and holography},'' {\em
  Adv.Theor.Math.Phys.} {\bfseries 2} (1998) 253--291,
\href{http://arxiv.org/abs/hep-th/9802150}{{\ttfamily arXiv:hep-th/9802150
  [hep-th]}}.

\bibitem{Witten:2007kt}
E.~Witten, ``{Three-Dimensional Gravity Revisited},''
\href{http://arxiv.org/abs/0706.3359}{{\ttfamily arXiv:0706.3359 [hep-th]}}.

\bibitem{Maloney:2007ud}
A.~Maloney and E.~Witten, ``{Quantum Gravity Partition Functions in Three
  Dimensions},'' {\em JHEP} {\bfseries 1002} (2010) 029,
\href{http://arxiv.org/abs/0712.0155}{{\ttfamily arXiv:0712.0155 [hep-th]}}.

\bibitem{Maldacena:1998bw}
J.~M. Maldacena and A.~Strominger, ``{AdS(3) black holes and a stringy
  exclusion principle},'' {\em JHEP} {\bfseries 9812} (1998) 005,
\href{http://arxiv.org/abs/hep-th/9804085}{{\ttfamily arXiv:hep-th/9804085
  [hep-th]}}.

\bibitem{Keller:2014xba}
C.~A. Keller and A.~Maloney, ``{Poincare Series, 3D Gravity and CFT
  Spectroscopy},'' {\em JHEP} {\bfseries 1502} (2015) 080,
\href{http://arxiv.org/abs/1407.6008}{{\ttfamily arXiv:1407.6008 [hep-th]}}.

\bibitem{Kim:2014bga}
J.~Kim and M.~Porrati, ``{Long String Dynamics in Pure Gravity on AdS$_3$},''
\href{http://arxiv.org/abs/1410.3424}{{\ttfamily arXiv:1410.3424 [hep-th]}}.

\bibitem{Kim:2015bba}
J.~Kim and M.~Porrati, ``{More on Long String Dynamics in Gravity on AdS3:
  Spinning Strings and Rotating BTZ},''
\href{http://arxiv.org/abs/1503.06875}{{\ttfamily arXiv:1503.06875 [hep-th]}}.

\bibitem{Achucarro:1987vz}
A.~Achucarro and P.~Townsend, ``{A Chern-Simons Action for Three-Dimensional
  anti-De Sitter Supergravity Theories},''
{\em Phys.Lett.} {\bfseries B180} (1986) 89.

\bibitem{Witten:1988hc}
E.~Witten, ``{(2+1)-Dimensional Gravity as an Exactly Soluble System},''
{\em Nucl.Phys.} {\bfseries B311} (1988) 46.

\bibitem{Mansson:2000sj}
T.~Mansson and B.~Sundborg, ``{Multi - black hole sectors of AdS(3) gravity},''
  {\em Phys.Rev.} {\bfseries D65} (2002) 024025,
\href{http://arxiv.org/abs/hep-th/0010083}{{\ttfamily arXiv:hep-th/0010083
  [hep-th]}}.

\bibitem{Castro:2011iw}
A.~Castro, R.~Gopakumar, M.~Gutperle, and J.~Raeymaekers, ``{Conical Defects in
  Higher Spin Theories},'' {\em JHEP} {\bfseries 1202} (2012) 096,
\href{http://arxiv.org/abs/1111.3381}{{\ttfamily arXiv:1111.3381 [hep-th]}}.

\bibitem{Cardy:1986ie}
J.~L. Cardy, ``{Operator Content of Two-Dimensional Conformally Invariant
  Theories},''
{\em Nucl.Phys.} {\bfseries B270} (1986) 186--204.

\bibitem{Strominger:1997eq}
A.~Strominger, ``{Black hole entropy from near horizon microstates},'' {\em
  JHEP} {\bfseries 9802} (1998) 009,
\href{http://arxiv.org/abs/hep-th/9712251}{{\ttfamily arXiv:hep-th/9712251
  [hep-th]}}.

\bibitem{Carlip:1995qv}
S.~Carlip, ``{The (2+1)-Dimensional black hole},'' {\em Class.Quant.Grav.}
  {\bfseries 12} (1995) 2853--2880,
\href{http://arxiv.org/abs/gr-qc/9506079}{{\ttfamily arXiv:gr-qc/9506079
  [gr-qc]}}.

\bibitem{Compere:2008us}
G.~Comp{\`e}re and D.~Marolf, ``{Setting the boundary free in AdS/CFT},'' {\em
  Class.Quant.Grav.} {\bfseries 25} (2008) 195014,
\href{http://arxiv.org/abs/0805.1902}{{\ttfamily arXiv:0805.1902 [hep-th]}}.

\bibitem{Compere:2013bya}
G.~Comp{\`e}re, W.~Song, and A.~Strominger, ``{New Boundary Conditions for
  $AdS_3$},'' {\em JHEP} {\bfseries 1305} (2013) 152,
\href{http://arxiv.org/abs/1303.2662}{{\ttfamily arXiv:1303.2662 [hep-th]}}.

\bibitem{Troessaert:2013fma}
C.~Troessaert, ``{Enhanced asymptotic symmetry algebra of $AdS$$_{3}$},'' {\em
  JHEP} {\bfseries 1308} (2013) 044,
\href{http://arxiv.org/abs/1303.3296}{{\ttfamily arXiv:1303.3296 [hep-th]}}.

\bibitem{Avery:2013dja}
S.~G. Avery, R.~R. Poojary, and N.~V. Suryanarayana, ``{An sl(2,$\mathbb{R}$)
  current algebra from $AdS_3$ gravity},'' {\em JHEP} {\bfseries 1401} (2014)
  144,
\href{http://arxiv.org/abs/1304.4252}{{\ttfamily arXiv:1304.4252 [hep-th]}}.

\bibitem{Apolo:2014tua}
L.~Apolo and M.~Porrati, ``{Free boundary conditions and the AdS$_3$/CFT$_2$
  correspondence},'' {\em JHEP} {\bfseries 1403} (2014) 116,
\href{http://arxiv.org/abs/1401.1197}{{\ttfamily arXiv:1401.1197 [hep-th]}}.

\bibitem{Balasubramanian:1999re}
V.~Balasubramanian and P.~Kraus, ``{A Stress tensor for Anti-de Sitter
  gravity},'' {\em Commun.Math.Phys.} {\bfseries 208} (1999) 413--428,
\href{http://arxiv.org/abs/hep-th/9902121}{{\ttfamily arXiv:hep-th/9902121
  [hep-th]}}.

\bibitem{Arcioni:2002vv}
G.~Arcioni, M.~Blau, and M.~O'Loughlin, ``{On the boundary dynamics of
  Chern-Simons gravity},'' {\em JHEP} {\bfseries 0301} (2003) 067,
\href{http://arxiv.org/abs/hep-th/0210089}{{\ttfamily arXiv:hep-th/0210089
  [hep-th]}}.

\bibitem{Balog:1988jb}
J.~Balog, L.~O'Raifeartaigh, P.~Forg\'acs, and A.~Wipf, ``{Consistency of
  String Propagation on Curved Space-Times: An SU(1,1) Based Counterexample},''
{\em Nucl.Phys.} {\bfseries B325} (1989) 225.

\bibitem{Hwang:1990aq}
S.~Hwang, ``{No ghost theorem for SU(1,1) string theories},''
{\em Nucl.Phys.} {\bfseries B354} (1991) 100--112.

\bibitem{Henningson:1991jc}
M.~Henningson, S.~Hwang, P.~Roberts, and B.~Sundborg, ``{Modular invariance of
  SU(1,1) strings},''
{\em Phys.Lett.} {\bfseries B267} (1991) 350--355.

\bibitem{Evans:1998qu}
J.~M. Evans, M.~R. Gaberdiel, and M.~J. Perry, ``{The no ghost theorem for
  AdS(3) and the stringy exclusion principle},'' {\em Nucl.Phys.} {\bfseries
  B535} (1998) 152--170,
\href{http://arxiv.org/abs/hep-th/9806024}{{\ttfamily arXiv:hep-th/9806024
  [hep-th]}}.

\bibitem{Giveon:1998ns}
A.~Giveon, D.~Kutasov, and N.~Seiberg, ``{Comments on string theory on
  AdS(3)},'' {\em Adv.Theor.Math.Phys.} {\bfseries 2} (1998) 733--780,
\href{http://arxiv.org/abs/hep-th/9806194}{{\ttfamily arXiv:hep-th/9806194
  [hep-th]}}.

\bibitem{deBoer:1998pp}
J.~de~Boer, H.~Ooguri, H.~Robins, and J.~Tannenhauser, ``{String theory on
  AdS(3)},'' {\em JHEP} {\bfseries 9812} (1998) 026,
\href{http://arxiv.org/abs/hep-th/9812046}{{\ttfamily arXiv:hep-th/9812046
  [hep-th]}}.

\bibitem{Kutasov:1999xu}
D.~Kutasov and N.~Seiberg, ``{More comments on string theory on AdS(3)},'' {\em
  JHEP} {\bfseries 9904} (1999) 008,
\href{http://arxiv.org/abs/hep-th/9903219}{{\ttfamily arXiv:hep-th/9903219
  [hep-th]}}.

\bibitem{Maldacena:2000hw}
J.~M. Maldacena and H.~Ooguri, ``{Strings in AdS(3) and SL(2,R) WZW model 1.:
  The Spectrum},'' {\em J.Math.Phys.} {\bfseries 42} (2001) 2929--2960,
\href{http://arxiv.org/abs/hep-th/0001053}{{\ttfamily arXiv:hep-th/0001053
  [hep-th]}}.

\bibitem{Maldacena:2000kv}
J.~M. Maldacena, H.~Ooguri, and J.~Son, ``{Strings in AdS(3) and the SL(2,R)
  WZW model. Part 2. Euclidean black hole},'' {\em J.Math.Phys.} {\bfseries 42}
  (2001) 2961--2977,
\href{http://arxiv.org/abs/hep-th/0005183}{{\ttfamily arXiv:hep-th/0005183
  [hep-th]}}.

\bibitem{Maldacena:2001km}
J.~M. Maldacena and H.~Ooguri, ``{Strings in AdS(3) and the SL(2,R) WZW model.
  Part 3. Correlation functions},'' {\em Phys.Rev.} {\bfseries D65} (2002)
  106006,
\href{http://arxiv.org/abs/hep-th/0111180}{{\ttfamily arXiv:hep-th/0111180
  [hep-th]}}.

\bibitem{Coussaert:1995zp}
O.~Coussaert, M.~Henneaux, and P.~van Driel, ``{The Asymptotic dynamics of
  three-dimensional Einstein gravity with a negative cosmological constant},''
  {\em Class.Quant.Grav.} {\bfseries 12} (1995) 2961--2966,
\href{http://arxiv.org/abs/gr-qc/9506019}{{\ttfamily arXiv:gr-qc/9506019
  [gr-qc]}}.

\bibitem{Banados:1999gh}
M.~Ba\~nados and A.~Ritz, ``{A Note on classical string dynamics on AdS(3)},''
  {\em Phys.Rev.} {\bfseries D60} (1999) 126004,
\href{http://arxiv.org/abs/hep-th/9906191}{{\ttfamily arXiv:hep-th/9906191
  [hep-th]}}.

\bibitem{Sundborg:2013bya}
B.~Sundborg, ``{Mapping pure gravity to strings in three-dimensional anti-de
  Sitter geometry},''
\href{http://arxiv.org/abs/1305.7470}{{\ttfamily arXiv:1305.7470 [hep-th]}}.

\bibitem{Brown:1992br}
J.~D. Brown and J.~W. York, Jr., ``{Quasilocal energy and conserved charges
  derived from the gravitational action},'' {\em Phys.Rev.} {\bfseries D47}
  (1993) 1407--1419,
\href{http://arxiv.org/abs/gr-qc/9209012}{{\ttfamily arXiv:gr-qc/9209012
  [gr-qc]}}.

\bibitem{Detournay:2014fva}
S.~Detournay, D.~Grumiller, F.~Sch{\"o}ller, and J.~Simón, ``{Variational
  principle and one-point functions in three-dimensional flat space Einstein
  gravity},'' {\em Phys.Rev.} {\bfseries D89} no.~8, (2014) 084061,
\href{http://arxiv.org/abs/1402.3687}{{\ttfamily arXiv:1402.3687 [hep-th]}}.

\bibitem{Miskovic:2006tm}
O.~Miskovic and R.~Olea, ``{On boundary conditions in three-dimensional AdS
  gravity},'' {\em Phys.Lett.} {\bfseries B640} (2006) 101--107,
\href{http://arxiv.org/abs/hep-th/0603092}{{\ttfamily arXiv:hep-th/0603092
  [hep-th]}}.

\bibitem{Henningson:1998gx}
M.~Henningson and K.~Skenderis, ``{The Holographic Weyl anomaly},'' {\em JHEP}
  {\bfseries 9807} (1998) 023,
\href{http://arxiv.org/abs/hep-th/9806087}{{\ttfamily arXiv:hep-th/9806087
  [hep-th]}}.

\bibitem{Skenderis:1999nb}
K.~Skenderis and S.~N. Solodukhin, ``{Quantum effective action from the AdS /
  CFT correspondence},'' {\em Phys.Lett.} {\bfseries B472} (2000) 316--322,
\href{http://arxiv.org/abs/hep-th/9910023}{{\ttfamily arXiv:hep-th/9910023
  [hep-th]}}.

\bibitem{Fefferman:1985ok}
C.~Fefferman and C.~R. Graham, ``Conformal invariants,'' in {\em Elie Cartan et
  les Math\'ematiques d'aujourd'hui}.
\newblock Asterisque, 1985.

\bibitem{Banados:1998gg}
M.~Ba\~nados, ``{Three-dimensional quantum geometry and black holes},'' {\em
  AIP Conf.Proc.} {\bfseries 484} (1999) 147--169,
\href{http://arxiv.org/abs/hep-th/9901148}{{\ttfamily arXiv:hep-th/9901148
  [hep-th]}}.

\bibitem{Moore:1989yh}
G.~W. Moore and N.~Seiberg, ``{Taming the Conformal Zoo},''
{\em Phys.Lett.} {\bfseries B220} (1989) 422.

\bibitem{Elitzur:1989nr}
S.~Elitzur, G.~W. Moore, A.~Schwimmer, and N.~Seiberg, ``{Remarks on the
  Canonical Quantization of the Chern-Simons-Witten Theory},''
{\em Nucl.Phys.} {\bfseries B326} (1989) 108.

\bibitem{Polyakov:1984et}
A.~M. Polyakov and P.~Wiegmann, ``{Goldstone Fields in Two-Dimensions with
  Multivalued Actions},''
{\em Phys.Lett.} {\bfseries B141} (1984) 223--228.

\end{thebibliography}\endgroup
\end{document}